\documentclass[12pt,preprint]{aastex}
\usepackage{lscape}

\begin{document}

\title{The Fermi/LAT sky as seen by INTEGRAL/IBIS}

\author{Ubertini P.\altaffilmark{1}, Sguera V.\altaffilmark{1,2},
Stephen J.B.\altaffilmark{2}, Bassani L.\altaffilmark{2}, Bazzano A.\altaffilmark{1},
Bird A.J.\altaffilmark{3}}

\altaffiltext{1}{INAF/IASF, Rome, Italy}
\altaffiltext{2}{INAF/IASF, Bologna, Italy}
\altaffiltext{3}{School of Physics and Astronomy, University of
Southampton, SO17 1BJ, UK}

\begin{abstract}
In this letter we present the result of the cross correlation
between the 4th INTEGRAL/IBIS soft gamma-ray catalog, in the range
20-100 keV, and the Fermi LAT bright source list  of objects
emitting in the 100 MeV - 100 GeV range. The main result is that
only a minuscule part of the more than 720 sources detected by
INTEGRAL and the population of 205 Fermi LAT sources are detected
in both spectral regimes. This is in spite of the mCrab INTEGRAL
sensitivity for both galactic and extragalactic sources and the
breakthrough, in terms of sensitivity, achieved by Fermi at
MeV-GeV energies. The majority of the 14 Fermi LAT sources clearly
detected in the 4th INTEGRAL/IBIS catalog are optically identified
AGNs (10) complemented by 2 isolated pulsars (Crab and Vela) and 2
High Mass X-Ray Binaries (HMXB, LS I +61$^{\circ}$303 and LS
5039). Two more possible associations have been found: one is 0FGL
J1045.6-5937, possibly the counterpart at high energy of the
massive colliding wind binary system Eta Carinae, discovered to be
a soft gamma ray emitter by recent INTEGRAL observations and 0FGL
J1746.0-2900 coincident with IGR J17459-2902, but still not
identified with any known object at lower energy. For the
remaining 189 Fermi LAT sources no INTEGRAL counterpart was found
and we report the 2$\sigma$ upper limit in the energy band 20-40
keV.

\end{abstract}

\section{Introduction}
In the last few
years, our knowledge of the soft gamma-ray sky (E$>$20 keV) has
greatly improved thanks to the observations performed by the
imager IBIS (Ubertini et al. 2003) onboard INTEGRAL
(Winkler et al. 2003) and  by the BAT telescope (Barthelmy et al.
2005) onboard Swift (Gehrels et al. 2004). In particular, IBIS is
surveying the 20--100 keV sky with a sensitivity better than a
mCrab and a point source location accuracy of the order of a few
arcmin. The 4th IBIS catalog
has been just released, listing 723 soft gamma-ray sources (Bird et al.
2009). It represents an extension both in exposure and sky
coverage with respect to the previous catalog (Bird et al. 2007).
The IBIS soft gamma-ray  sky is surprisingly dynamic and diverse,
with many different types of
objects such as Active galactic Nuclei (AGNs), galactic X-ray
binaries, Cataclysmic Variables (CVs.) and also newly  discovered
classes of sources such as  Supergiant Fast X-ray Transients
(SFXTs, Sguera et al. 2005,2006) and highly absorbed supergiant
high mass X-ray binaries (Walter et al. 2006).

Recently, INTEGRAL and Swift have been joined in operation by the
Fermi gamma-ray satellite launched in June 2008, providing a
coverage of the sky over 10 orders of magnitude in energy, from
keV to GeV. Onboard this satellite, the Large Area Telescope (LAT,
Atwood et al. 2009) is an imaging high energy gamma-ray telescope
(20 MeV -- 300 GeV) which provides, with respect to its
predecessor EGRET, superior angular resolution, sensitivity, field
of view and observing efficiency. These improved
capabilities have allowed a catalog of  205 highly significant
detections ($>$10$\sigma$) after three months of an all
sky survey (Abdo et al. 2009a). Many Fermi LAT sources  have been
identified as AGNs (121) and  pulsars (30), but interestingly
there are also 2 High Mass X-ray Binaries (HMXBs) while the
remaining objects have no obvious counterparts. To identify those sources
IBIS is particularly suited since it provides a hard X/soft
gamma-ray point source location accuracy of the order of an
arcmin as well as information close to the
MeV band.

\section{Cross correlating the 4th IBIS catalog and the Fermi LAT bright source list}

Many of the detected Fermi LAT sources are expected to emit around
20 keV, i.e. in the IBIS regime. With the aim of investigating
such a link, we have cross-correlated the positions of the 205
Fermi LAT bright sources with the 4th IBIS catalog. Specifically,
following Stephen et al. (2005, 2006), we took both catalogs and
calculated the number of LAT sources which could be considered to
be coincident with IBIS sources as a function of distance. As a
control, we then created another database of 205 false sources
derived from the Fermi LAT catalogue but with positions mirrored
in both galactic longitude and latitude. The top panel of Fig. 1
shows the number
of coincidences for both the real and false datasets while the
bottom panel illustrates the difference between the two. The leveling off of the
 lower curve in Fig. 1 at a distance of about 4 arcminutes indicates that there is 
a real correlation, with $\sim$ 13-14 sources common to both catalogs. In comparison, 
at this distance only one source in the fake dataset appears to be correlated.
 The 14 associations are listed in table 1 (the first 14 sources in the list). The most likely chance correlation is 
0FGL J1746.0-2900/IGR J17459-2902 found in the galactic center region, 
i.e. where a likelihood of an association by chance is very high due to INTEGRAL 
reaching the confusion limit and the presence of strong Fermi diffuse emission
 overlapping with compact sources.

In applying this technique, we only employ the positions of the sources, and
 not the errors on these positions and therefore sources which have relatively 
large error boxes could become "lost" in the chance association distribution. In 
order to make use of this extra information and retrieve these associations we 
have applied another procedure in which we compare the distance between 
IBIS and LAT sources with the quadratic sum of their corresponding error radii. 
We first convert the Fermi 95\% and IBIS 90\% quoted error radii to the 
equivalent 1σ values, which were then added in quadrature. This value is 
then compared with the distance between the Fermi and IBIS catalog positions 
and sources are assumed to be potentially correlated when their separation is less 
than 2.15$\sigma$, corresponding to 90\% for a 2-D normal distribution. The result of 
this second procedure is to reaffirm the 14 associations found using the first method 
and to recuperate 2 more sources (the last two rows in table 1). 
The first is 1ES 0033+59.5 which appears in both the IBIS and Fermi catalogs 
and so is assumed to be a correct association, while the second corresponds to 
Eta Carinae/0FGL J1045.6-5937, listed in the Fermi LAT bright source catalog as 
unidentiﬁed.

 Fig. 2 shows its 95\%
error circle ($\sim$ 7.4 arcmin radius) superimposed on the IBIS
18-60 keV significance mosaic with an exposure of $\sim$ 2.3 Ms.
The clear IBIS detection ($\sim$ 5.5$\sigma$) is Eta Carinae,
located right on the border of the LAT error circle. It is worth
pointing out that the AGILE gamma-ray satellite  detected an MeV
source (1AGL J1043-5931, 95\% error radius of 0.68 degrees)
consistent with the position of Eta  Carinae (Tavani et al.,
2009). For all the remaining 189 Fermi LAT sources for which no
IBIS counterpart was found, we calculated the 2$\sigma$ upper
limit in the energy band 20-40 keV, and these are listed in Table
2.

\section{Discussion}

The Fermi sources firmly detected by IBIS comprise
the 10 optically identified AGNs complemented by the 2 isolated
pulsars Crab and Vela and 2 HMXB, while there
are a further two possible associations with unidentified Fermi
sources. This can be interpreted as an indication that the
acceleration processes dominating in the high energy range are
only partially efficient in the soft gamma ray regime. Vice versa,
the well populated INTEGRAL sky, comprising 13\% LMXB, 13\% HMXB,
35\% AGN, 10\% Pulsar, SNR and CV and 29\%
Unknown emitter, is barely detectable at MeV energies, in spite of the unprecedented sensitivity of Fermi. Figure 3 shows the gamma-ray flux (100 MeV - 1 GeV) of each catalogued LAT source as a function of the IBIS flux in the 20-40 keV range. First of all we note that no LAT sources (except for the peculiar object ETA Carinae) have been detected by IBIS at a flux below 1 mCrab despite the fact that a large fraction
of the IBIS sky has a detection limit well below this value (typically about 0.1 mCrab) and is well populated with hard X-ray sources. This indicates that the typical LAT source is either very strong or very weak in the IBIS energy range providing important information on the Spectral Energy Distribution (SED) of these objects. Further, only a couple of the Seyfert 1 and 2 detected by INTEGRAL (Bassani, et al., 2009, Beckmann, et al., 2009, Molina, et al., 2009) are present in the Fermi catalog, i.e. the two bright radio galaxies NGC1275 and Cen A. In contrast to blazars, most radio galaxies have large inclination angles and hence their emission is not significantly amplified due to Doppler beaming; however, if the jet has velocity gradients (Georganopoulos \& Kazanas, 2003, Ghisellini et al., 2005, Lenain et al., 2008) then it is possible to produce bright gamma-ray emission via the IC process where one component sees the (beamed) radiation produced by the other, and this enhances the IC emission of all components in the jet.

The majority of the associations reported in Table 1 refer to
blazars, mainly detected by IBIS during TOO observations, hence in
flaring, or in deep studies with long devoted exposures or because
they are bright objects over the entire gamma-ray band (see Fig.3).
Their detection in the IBIS band is therefore not unexpected
and suggests that the number of blazars common to both gamma-ray
regimes  will increase as the INTEGRAL and Fermi catalogues
expand. Interestingly, IBIS also sees both types of blazars, i.e.
Flat Spectrum Radio Quasars (FLSQ) and  BL Lac type objects. In
the widely adopted scenario of blazars, a single population of
high-energy electrons in a relativistic jet radiates from the
radio/FIR to the UV/soft X-rays by the synchrotron process and at
higher frequencies by inverse Compton scattering of soft photons
present either in the jet (Synchrotron Self-Compton [SSC] model),
in the surrounding material (External Compton [EC] model), or in
both (Ghisellini et al. 1998 and references therein). Therefore, a
strong signature of the Blazar nature of a source is a double
peaked structure in the SED, with the synchrotron component peaking anywhere from Infrared to X-rays and the inverse Compton extending up to GeV or even TeV gamma-rays (Maraschi and Tavecchio, 2003 Sambruna et al., 2006).

The analysis of the gamma-ray properties of Fermi blazars  (Abdo
et al. 2009) have revealed that the average GeV spectra of BL Lac
objects are significantly harder ($\Gamma$ $\sim$1.99$\pm$0.22)
than the spectra of FSRQs ($\Gamma$ $\sim$ 2.40 $\pm$0.17). This
is probably  due to Fermi observing different parts of the Compton
bump in the different blazar populations: the ascending part in BL
Lac and the descending part in FSRQ. A similar analysis can be
performed with IBIS detected blazars despite the poor statistics.
The class of BL Lac objects seen by IBIS, including BL Lac, Mkn501
and Mkn421, are weak emitters in soft gamma rays when not flaring
and have 20-100 keV spectral indices usually steeper than 2
(Guetta et al., 2004, Lichti et al., 2008). Conversely, the FSRQ,
including 3C454.3, PKS1830-211, 3C273 and 3C279 are characterised
by harder spectra, though more easily detectable by IBIS in the
20-100 keV range even if in the quiescent state (De Rosa et al,
2005, Vercellone, et al., 2008, Iafrate et al. 2009, Chernyakova
et al 2007). Their sustained emission in this energy range is due
to IC plus EC on a Broad Line Region (BLR) or disk photons in the
case of 3C454.3,(Vercellone et al., 2008), to a mix of SSC plus EC
from a BLR and torus for PKS 1830-211, (De Rosa et al, 2005) and
to a more complex scenario in the cases of 3C279 and 3C273
(Costamante, et al., 2009, S. Soldi et al.,2008). The difference
in the spectral shapes of BL Lac and FSRQ recently reported by
INTEGRAL/IBIS and Swift/BAT (Ajello et al. 2009) indicates that
their energy range probes the end part of the Synchrotron peak in BL Lacs and the ascending part of the Compton peak in FSRQ.

As far as galactic sources are concerned, the only two Fermi
isolated pulsars detected by IBIS (Crab and Vela) are among the
brightest gamma-ray sources known (100 MeV - 1 GeV). IBIS
also detected a few more pulsars including the newly discovered
hard X-ray pulsars PSR B1509-58 and PSR B0540-69 as well as  some
radio quiet/dim young pulsars located in SNRs (PSR BJ1846-0258,
PSR B1811-1925 and PSR B1617-5055). These spin-powered pulsars are
characterized by a hard spectral tail and different pulse profile
though the radio quiet/dim ones show both light curves  and
spectral shape  of the pulsed spectrum similar to PSR B1509-58.
The non-thermal hard X-ray component can be explained with
synchrotron emission both in a slot-gap or outer-gap model
(Hirotani et al. 2008), but there is not yet any firm conclusion
on the origin of the high energy electrons and how and
where they  are being accelerated. Apart from the Crab with a
luminosity peak at around 100 keV, all the remaining gamma-ray
pulsars show a dominating emission at energies higher than 10 MeV.

Of particular interest are the IBIS detections of the two Fermi
HMXBs LS 5309 and LS I +61 303. It has been suggested that these
are both microquasars although another scenario is that they host
young non-accreting pulsars and are powered by the interaction
between the pulsars' relativistic wind and the wind from the
massive stellar companion (Dubus 2006).  Both these HMXBs are
those with the most exposure (apart from the Crab) at TeV
energies, and now also in the MeV/GeV regime. The measurements
indicate that the emission mechanisms active in these two energy
ranges are not correlated (Abdo et al. 2009b, Holder et al. 2009).
INTEGRAL has detected several (10-15) BHC/microquasar candidates,
but it is necessary that Fermi detects more of these sources in
order to be able to disentangle the emission mechanisms active at
high energies and at soft gamma-rays in these objects.

\section{Conclusions}

From our analysis it appears that MeV/GeV Fermi sources are not
commonly associated with  IBIS sources in the range 20 -- 100 keV.
The handful of objects common to both surveys comprise so far
mainly FSRQ and BL Lac, with no XRB apart from the
two microquasars. Equally absent are the AXP which are strong
emitters in the  keV-MeV range with a total energy rising in
$\nu$F$\nu$ and no cut-off detected up to a few hundreds of keV
(Kuiper and Hermsen 2009), implying some kind of switch-off
mechanism in the MeV regime. Similarly, SGR and Magnetars,
detected even in quiescence mode by IBIS (Rea et al, 2009) and
among the brightest sources when flaring (Kouveliotou et al. 1999,
Gotz et al. 2006, Israel et al. 2008) are not detected so far by
Fermi/LAT.

\begin{landscape}
\begin{table*}[t!]
\begin{center}
\caption {Spatial correlation between Fermi LAT and IBIS
sources.\tablenotemark{1}}
\begin{tabular}{cccccccc}
\hline \hline
Fermi        & RA      & Dec     & IBIS           & IBIS         & distance   &                                                               & ID source \\
(0FGL)       & (deg)   &(deg)    & RA (deg)             & Dec  (deg)         &   (arcmin) & $\frac{distance}{\sqrt{Fermi R_{1\sigma}^2 + Ibis R_{1\sigma}^2}}$  &          \\
\hline
J1229.1+0202 & 187.287 & 2.045   & 187.279        & 2.049        & 0.533      & 0.258        & 3C 273\\
J1325.4-4303 & 201.353 & -43.062 & 201.363        & -43.021      & 2.5        & 0.332       &  Cen A \\
J1653.9+3946 & 253.492 & 39.767  & 253.488        & 39.753       & 0.85       & 0.410     & Mkn 501\\
J2202.4+4217 & 330.662 & 42.299  & 330.677        & 42.293       & 2.01       & 0.481  & BL Lac\\
J2254.0+1609 & 343.502 & 16.151  & 343.489        & 16.149       & 0.75       & 0.558    & 3C 454.3\\
J0240.3+6113 & 40.093  & 61.225  &  40.119     & 61.240       & 1.16       & 0.562  & LS I+616 303\\
J0320.0+4131 & 50.000  & 41.524  & 49.967      & 41.532       & 1.55       & 0.645   &  NGC 1275 \\
J1256.1-0547 & 194.034 & -5.8   & 194.044     & -5.770       & 1.88       & 0.827 &  3C 279 \\
J1104.5+3811 & 166.137 & 38.187 & 166.119     & 38.207       & 1.46       & 1.067   & Mkn 421\\
J1833.4-2106 & 278.37  & 21.103 & 278.415     & -21.063      & 3.46       & 1.180  & PKS 1830-211\\
J0534.6+2201 & 83.653  & 22.022 & 83.629      & 22.017       & 1.36       & 1.190    & Crab \\
J1826.3-1451 & 276.595 & -14.86 & 276.525     & -14.847      & 4.13       & 1.406   &  LS 5039\\
J0835.4-4510 & 128.865 & -45.17 & 128.831 & -45.179       & 1.53       & 1.437  & Vela Pulsar\\
J1746.0-2900 & 266.506 &-29.005 & 266.485 &-29.043           & 2.51       & 1.472   & IGR J17459-2902\\
J0036.7+5951 & 9.177   & 59.854 & 8.964   & 59.83         & 6.56       & 1.802 & 1ES 0033+59.5 \\
J1045.6-5937 & 161.409 & -59.631& 161.206 & -59.704       & 7.5        & 2.073 &  Eta Carinae \\
\hline \hline
\end{tabular}
\end{center}
\tablenotetext{1}{Note: the sixth column is the distance between
the LAT and IBIS source positions while the seventh column is this
distance divided the square root of the quadratic sum of their
corresponding error radii.}

\end{table*}
\end{landscape}

\begin{figure}[t!]
\plotone{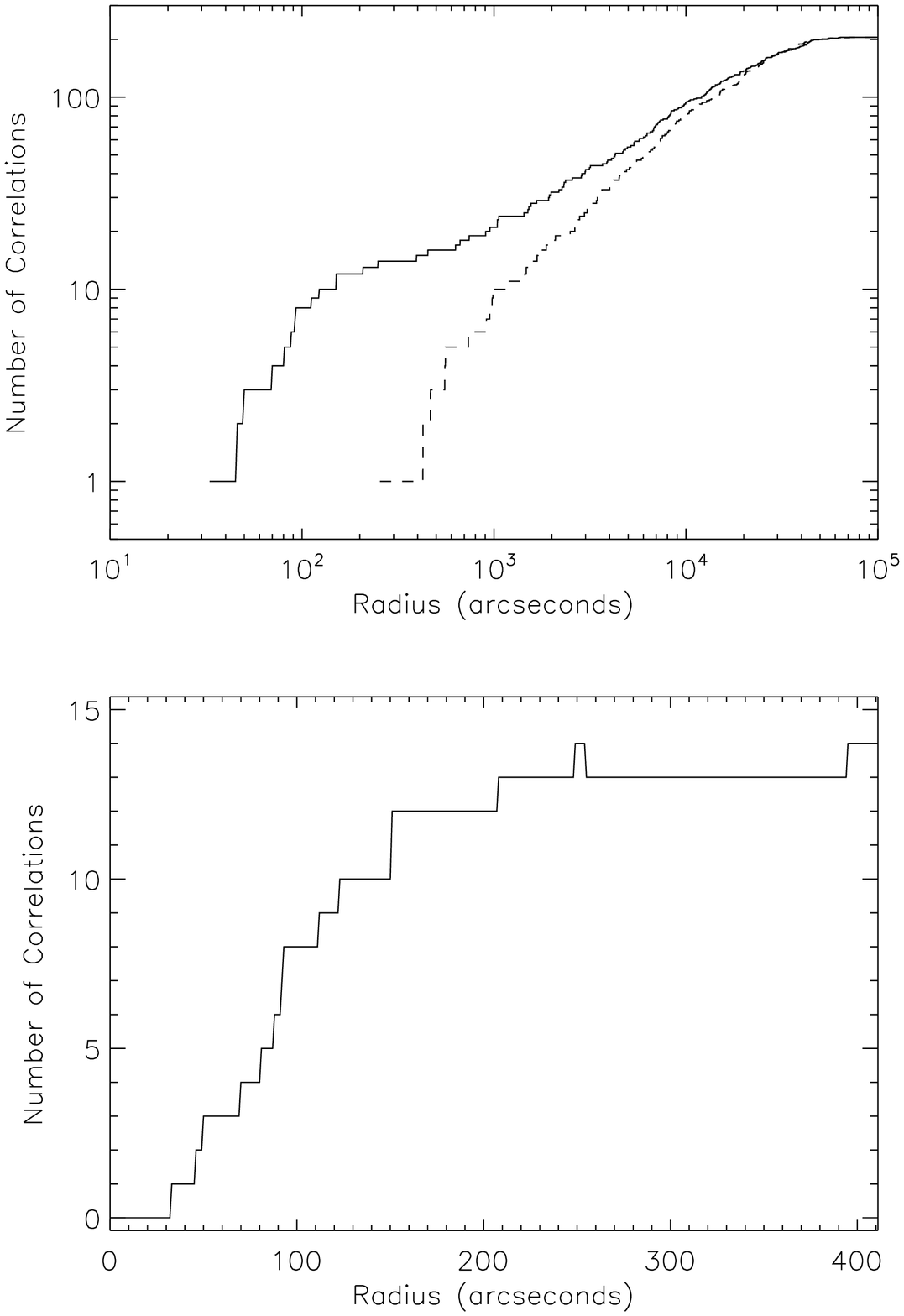} 
\caption{Top: the number of IBIS sources
associated with Fermi (solid line) and fake (dashed) objects. Bottom: the difference between the two upper curves revealing a correlation for a small number of sources:}
\end{figure}

\begin{figure}[t!]
\plotone{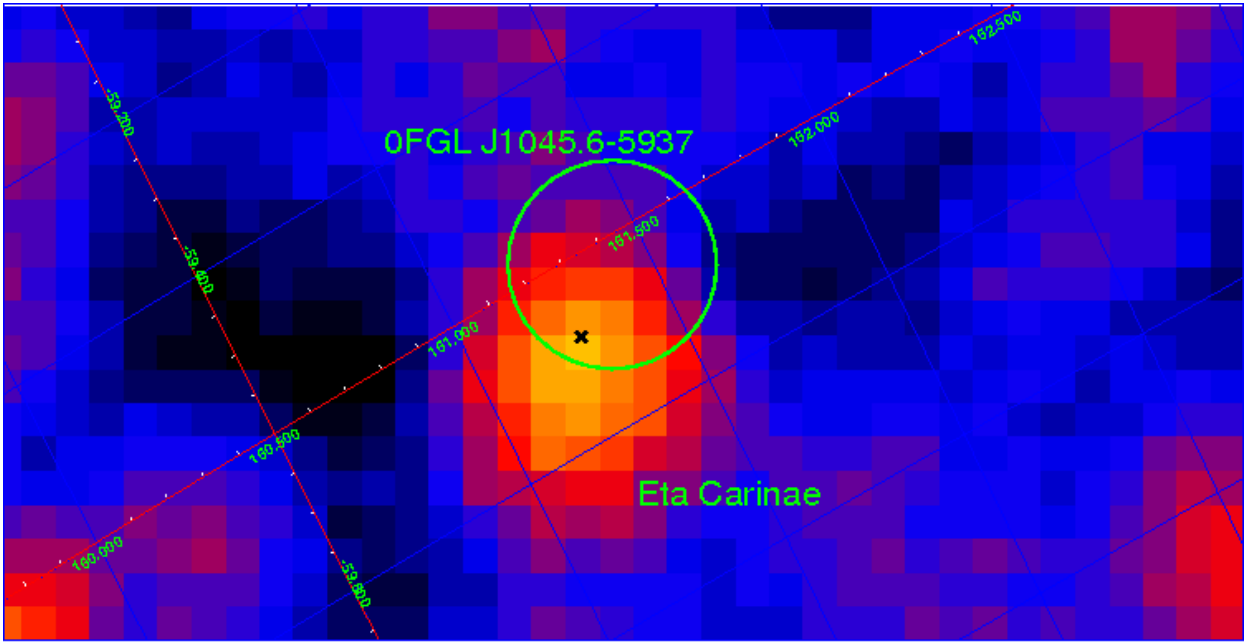}
\caption{IBIS 18-60 KeV significance map ($\sim$ 2.3 Ms exposure
time) superimposed on the Fermi LAT errror circle of the
unidentified source 0FGL J1045.6-5937. It is evident a clear IBIS
detection at $\sim$ 5.5$\sigma$. The optical position of Eta
Carinae is marked by a black cross point.}
\end{figure}

\begin{figure}[t!]
\plotone{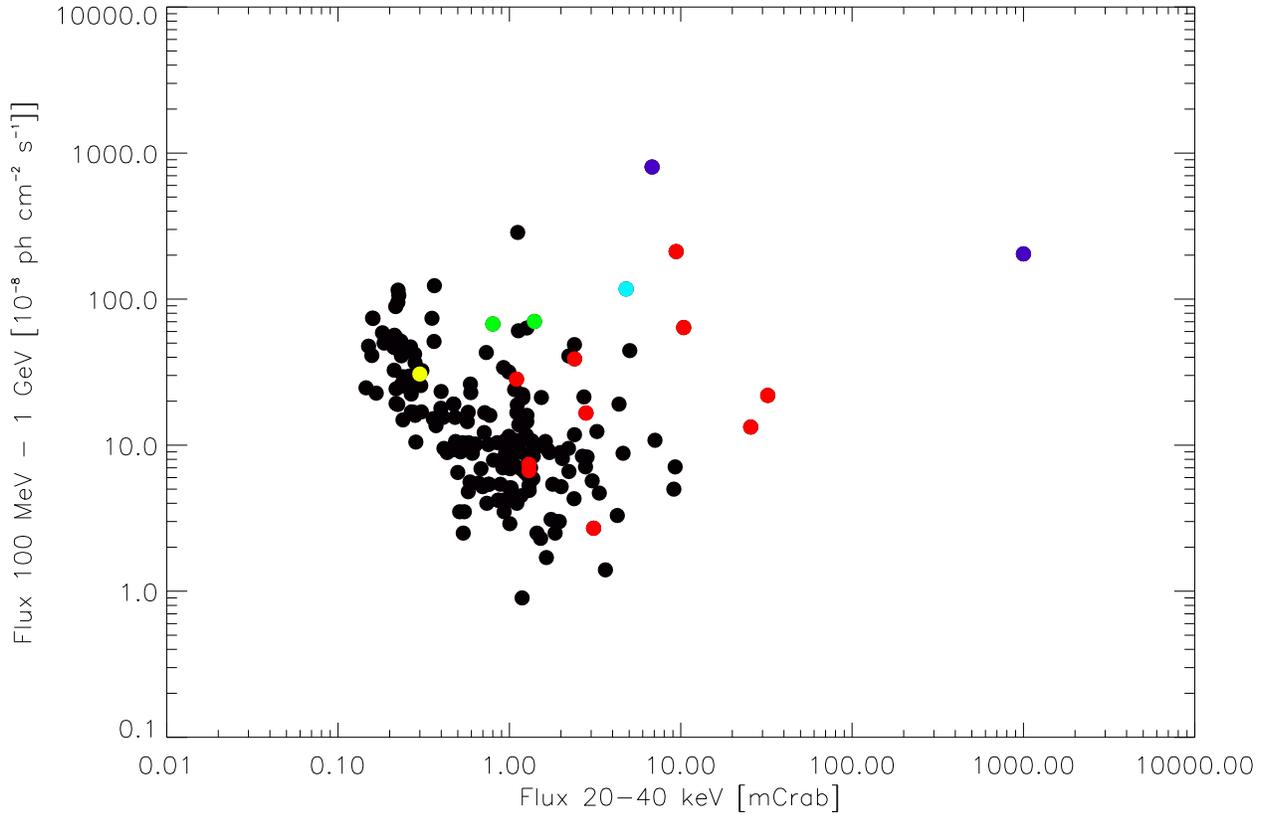}
\caption{Gamma-ray flux  (100 MeV - 1 GeV) of each Fermi  source
as a function of the corresponding 20--40 keV IBIS flux. The
coloured points refer to the IBIS detections, specifically red
points are blazars, dark blu are pulsars, green are HMXBs, yellow
is Eta Carinae and finally light blu is IGR J17459-2902. The black
points refer to IBIS non detection (2$\sigma$ upper limit). }
\end{figure}

\begin{landscape}
\begin{table*}
\begin{center}
\caption {2$\sigma$ upper limit (20-40 keV) for the 189 Fermi LAT
sources with no IBIS counterpart\tablenotemark{2}}
\begin{tabular}{cccccccc}
\hline \hline
     Fermi         &  upper limit    &     Fermi   &     upper limit   & Fermi  &  upper limit &    Fermi  &   upper limit     \\
     (0FGL)       &  20-40 KeV       &   (0FGL)   &   20-40 keV        & (0FGL)&   20-40 keV      &  (0FGL)   &  20-40 keV     \\
                   &  (mCrab)       &           &   (mCrab)        &          &   (mCrab)      &      &   (mCrab)       \\
\hline
        J0007.4+7303      &    0.3084       &     J0654.3+5042   &       2.3834       &     J1331.7-0506   &      0.5776      &  J1834.4-0841     &    0.2302   \\
        J0017.4-0503      &    2.3976       &    J0700.0-6611     &      0.5454          &      J1333.3+5058  &   0.6076     &      J1836.1-0727 &     0.2332    \\
\hline \hline
\end{tabular}
\end{center}
\tablenotetext{2}{Note. Table 2 is published in its entirety in
the electronic edition of the Astrophysical Journal Letters. A
portion is shown here for guidance regarding its form and content.
The first (and odd) column(s) list the Fermi LAT bright sources
emitting in the 100 MeV - 100 GeV range (Abdo et al., 2009); the
second (and even) column(s) report the Integral/IBIS 2sigma upper
limit in the energy band 20-40 keV. The IBIS fluxes are in
milliCrab = 7.6 x 10$^{-12}$erg cm$^{2}$s in the 20-40 KeV range.}
\end{table*}
\end{landscape}

\section{Acknowledgements}
Italian co-authors acknowledge ASI financial support via contract
I/08/07/0. The authors are grateful to Catia Spalletta for
professional editing of the manuscript.


\begin{thebibliography}{}

\bibitem{}Abdo, A. A., Ackerman, M., Ajello, M. et al 2009, ApJ 700, 597,2009
\bibitem{} Abdo, A. A., Ackermann, M., Ajello, M., et al. 2009a, ApJS, 183, 46
\bibitem{} Abdo, A. A., Ackerman, M., Ajello, M. et al. 2009b, ApJ letters in press, astro-ph 0907.4307
\bibitem{}Ajello, M., Costamante, L., Sambruna, M.R., et al. 2009, Ap.J 699, 603
\bibitem{} Atwood, W. B.; Abdo, A. A.; Ackermann, M., e al. 2009, ApJ, 697, 1071
\bibitem{} Bassani, L., Landi, R., Parisi, P., 2009, MNRAS, 395,L1
\bibitem{} Barthelmy, S.D. et al.; 2005, Space Science Reviews, 120,143
\bibitem{} Beckmann, V.; Soldi, S.; Ricci, C., et al. 2009, A\&A in press, astro-ph 0907.0654
\bibitem{} Bird, A. J.; Malizia, A.; Bazzano, A., et al. 2007,  ApJS,170,175
\bibitem{} Bird, A. J., Bazzano, A., Bassani, L., et al. 2009, ApJS, submitted
\bibitem{} M. Chernyakova, Neronov, A., Courvoisier, T. J.-L. et al. 2007, A\&A, 465,147
\bibitem{} Costamante,L., Aharonian,F., Buehler, R., Khangulyan, D., Reimer, A., Reimer, O.,arxiv.org/abs/0907.3966, 2009
\bibitem{} De Rosa, A, Piro, L., Tramacere, A., et al., 2005, A\&A, 438, 12
\bibitem{} Dubus, G. 2006, A\&A, 456, 801
\bibitem{} Gehrels, N.; Chincarini, G.; Giommi, P., et al. 2004, ApJ,611, 1005
\bibitem{} Georganopoulos, M. \& Kazanas, D. 2003b, ApJ, 594, L27
\bibitem{}Ghisellini G., Celotti A., Fossati G., et al.,1998, MNRAS, 301, 451
\bibitem{} Ghisellini,G., Tavecchio,  F., Chiaberge, M., 2005, A\&A, 432, 401
\bibitem{} Gotz D., S. Mereghetti, S. Molkov, et al., 2006, A\&A, 445, 313
\bibitem{} Guetta,D., Ghisellini, G., Lazzati, D., and A. Celotti, 2004, A\&A, 421, 877
\bibitem{} Holder, J. et al. 2009,  Proc 31st ICRC, Lodz, Poland, 2009, astroph 0907.3921
\bibitem{} Hirotani et al. 2008, ApJ, 688, L25
\bibitem{} Iafrate et al. 2009, Atel 2154
\bibitem{} Israel, G.L., P. Romano, V. Mangano, et al. 2008, ApJ, 685, 1114
\bibitem{} Kouveliotou C., T. Strohmayer, K. Hurley, et al. 1999, ApJL, 510, L115
\bibitem{} Kuiper, L. and Hermsen, W., 2009, A\&A, 501, 1031
\bibitem{} Lenain, J.-P., Boisson, C., Sol, et al. 2008, A\&A, 478, 111
\bibitem{} Lichti, G. G., Bottacini, E., Ajello, M., et al., 2008, A\&A, 486, 721
\bibitem{} Maraschi, L., Tavecchio, F., 2003, ApJ, 593, 667
\bibitem{} Molina, M.,  Bassani,. L.,  Malizia,.  A.,  et al., 2009, MNRAS in press, astro-ph 0906.2909
\bibitem{} Rea, N.; Israel, G. L.; Turolla, R., et al. 2009, MNRAS, 396, 2419
\bibitem{} Sambruna, R.M., Gliozzi, M., Tavecchio, F., 2006, ApJ, 652, 146
\bibitem{} Stephen, J. B.; Bassani, L.; Molina, M., et al. 2005, A\&A, 432L, 49
\bibitem{} Stephen, J. B.; Bassani, L.; Malizia, A., et al. 2006, A\&A, 445, 869
\bibitem{} Sguera, V.; Barlow, E. J.; Bird, A. J. et al. 2005, A\&A,444,221
\bibitem{} Sguera, V.; Bazzano, A.; Bird, A. J.,et al. 2006, ApJ, 646, 452
\bibitem{} Soldi, S., T\"{u}rler, M., Paltani, S., et al. A\&A, 2008, 486, 411
\bibitem{} Tavani, M.; Sabatini, S.; Pian, E., et al. 2009, ApJ, 698L, 142
\bibitem{} Ubertini, P.; Lebrun, F.; Di Cocco, G., et al. 2003, A\&A, 411L, 131
\bibitem{} Vercellone, S.; Chen, A. W.; Giuliani, et al.,  2008, ApJ, 676, L13
\bibitem{} Walter, R.; Zurita Heras, J.; Bassani, L., et al. 2006, A\&A, 453, 133
\bibitem{} Winkler, C.; Courvoisier, T. J.-L.; Di Cocco, G., e al. 2003, A\&A, 411L, 1



\end{thebibliography}
\end{document}